\documentclass[11pt]{article}
\usepackage{moriond,epsfig}

\def\be{\begin{equation}}
\def\ee{\end{equation}}
\def\bea{\begin{eqnarray}}
\def\eea{\end{eqnarray}}

\begin{document}

{\small \noindent IPPP/09/39\\
  MIT-CTP 4038 \\
   Saclay-IPhT-T09/059 \\
    SLAC--PUB--13630\\
     UCLA/09/TEP/48
}

\vspace*{2cm}
\title{Multi-jet cross sections at NLO with BlackHat and Sherpa}

\author{C.~F.~Berger${}^{a}$, Z.~Bern${}^b$,
L.~J.~Dixon${}^c$, F.~Febres Cordero${}^b$, D.~Forde${}^{c}$,
T.~Gleisberg${}^c$,  H. Ita${}^b$, 
D.~A.~Kosower${}^d$ and D.~Ma\^{\i}tre${}^e$}

\address{${}^a${Center for Theoretical Physics, Massachusetts Institute of Technology, Cambridge, MA 02139, USA} 
\\
${}^b$Department of Physics and Astronomy, UCLA, Los Angeles, CA 90095-1547, USA \\
${}^c$SLAC National Accelerator Laboratory, Stanford University,             Stanford, CA 94309, USA 
\\
${}^d$Institut de Physique Th\'eorique, CEA--Saclay,          F--91191 Gif-sur-Yvette cedex, France\\
${}^e$Department of Physics, University of Durham,          DH1 3LE, UK
}

\maketitle\abstracts{
In this talk, we report on a recent next-to-leading order QCD calculation of the production of a $W$ boson in association with three jets at hadron colliders. The computation is performed by combining two programs, {\tt BlackHat} for the computation of the virtual one-loop matrix elements and {\tt Sherpa} for the real emission part. The addition of NLO corrections greatly reduces the factorization and renormalization scale dependence of the theory prediction for this process. This result demonstrates the applicability of unitarity-based methods for hadron collider physics.}

\section{Introduction}

The production of a vector boson in association with jets is an important
process at the LHC. Apart from its interest as a test of QCD, it contributes
significantly to the background of many Standard Model processes ($t\bar t$
production, single top production, and Higgs decay to two vector bosons) and new
physics processes. Successful measurements of these processes require a reliable
theoretical description of the vector boson\,+\,jets processes. 

Leading-order QCD predictions for processes with jets suffer from a large
dependence on the renormalization and factorization scales. This problem
can be tamed by adding next-to-leading order (NLO) corrections.
Such corrections are composed of two parts. The real corrections to an
$n$-parton process arise when an additional parton is emitted, in an
$(n+1)$-parton process. One-loop $n$-parton amplitudes generate the 
virtual part of the NLO correction.




{\tt BlackHat}~\cite{BlackHatI} is a numerical implementation in C++ of so-called on-shell
methods for computing one-loop amplitudes.  The starting point for a
one-loop amplitude $A$ with massless propagators is its general decomposition 
in terms of scalar integrals,
\begin{equation}
A=\sum\limits_{i}c_4^iI_4^i+\sum\limits_{i}c_3^iI_3^i+\sum\limits_{i}c_2^i I_2^i  +R\quad,
\end{equation}
where $I_2^i$, $I_3^i$, $I_4^i$ are scalar bubble, triangle and box
integrals. The `rational term' $R$ is a rational function of 
spinor products and does not contain any logarithms. The objective of
on-shell methods is to determine the coefficients $c_n^i$ of the integrals
and $R$ without using Feynman diagrams. We refer the reader to the
literature~\cite{OnShellReview} for more details on these methods. 

In our numerical implementation, the coefficients of the integrals are
determined using the analytic approach of Forde~\cite{Forde}, which is
related to other recent methods~\cite{OPP}.  The rational term is computed 
using on-shell recursion relations for one-loop
amplitudes~\cite{Bootstrap}.  Numerical stability of the implementation
is achieved by using high-precision libraries~\cite{QD} when
(and only when) necessary. This stability has been demonstrated elsewhere~\cite{ICHEPBH}.
  

{\tt Sherpa}~\cite{Sherpa} is a C++ Monte Carlo event generator.
It can compute the real part of the NLO corrections in an automated
way~\cite{Amegic} using Catani and Seymour's dipole subtraction
method~\cite{CS}. In addition, the subtraction term integrated over
the unresolved phase space is provided. The results presented below have
been integrated over the relevant phase space using {\tt Sherpa}.


For the $W\,+\,$3~jets virtual cross section we used a leading-color (LC) approximation for the finite part of the virtual amplitude. 
This approximation amounts to neglecting the terms in the ratio of the 
virtual terms to the tree cross section that are suppressed by factors of 
$1/N_C^2$ (color suppressed) or $N_f/N_c$ (virtual quark loop). 
This approximation has been shown to be very good in the following section
for $W\,+\,$1,2~jets~\cite{W3jets}, so we expect it to be valid for 
$W\,+\,$3~jets.  A related, but different, approximation that includes
only a subset of partonic subprocesses has been used in another
computation~\cite{EMZ} of $W\,+\,$3~jets.  
The benefit of our approximation is that the number of (color ordered) 
primitive amplitudes to evaluate is significantly reduced.  
We checked agreement between the primitive amplitudes we used here
and those found in a different calculation~\cite{W3EGKMZ}.

\section{$\boldmath{W}$+3 jets at the Tevatron}\label{sec:Tevatron}

\begin{figure}
\includegraphics[scale=0.27]{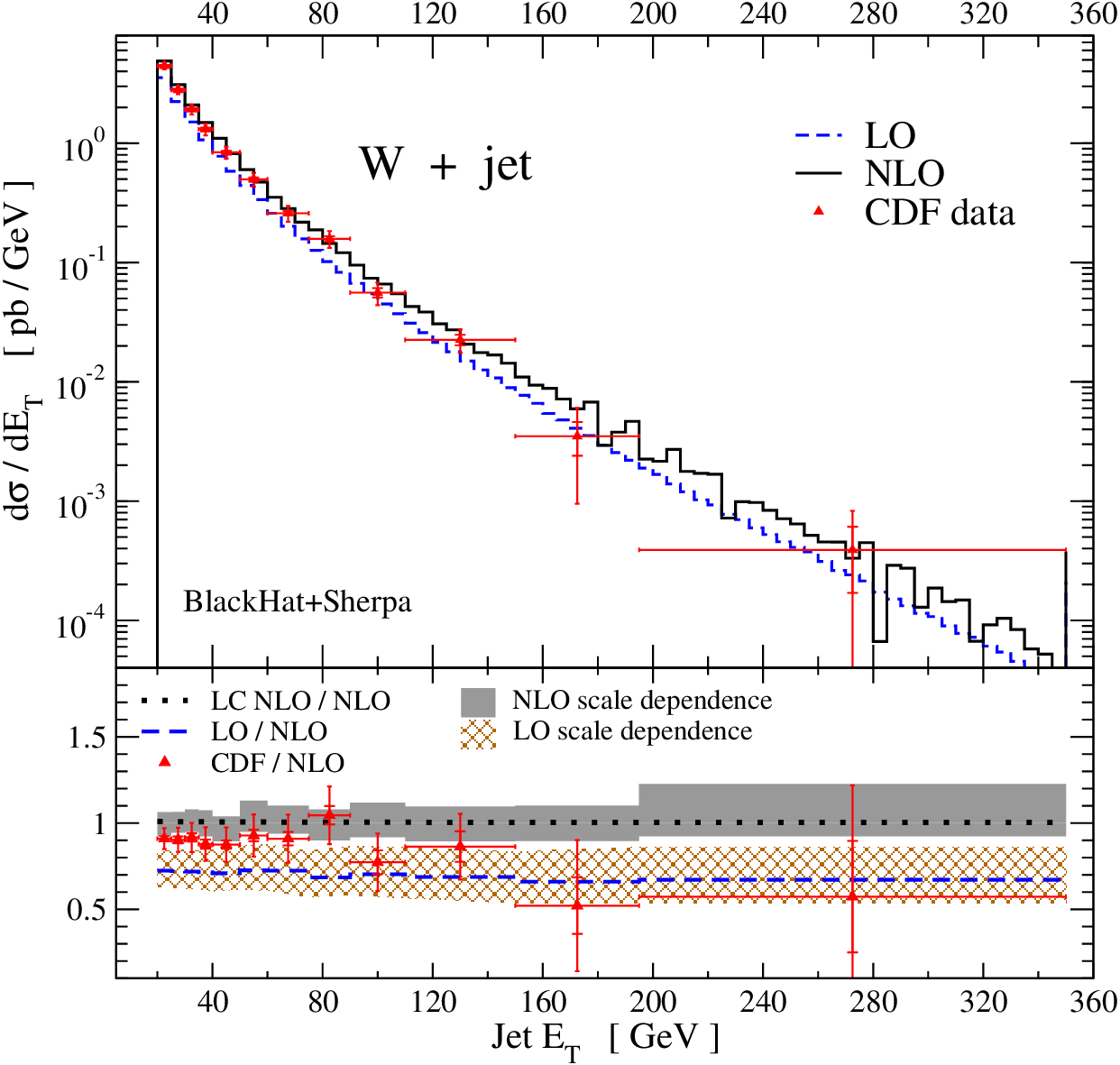}
\rule{1cm}{0cm}
\includegraphics[scale=0.27]{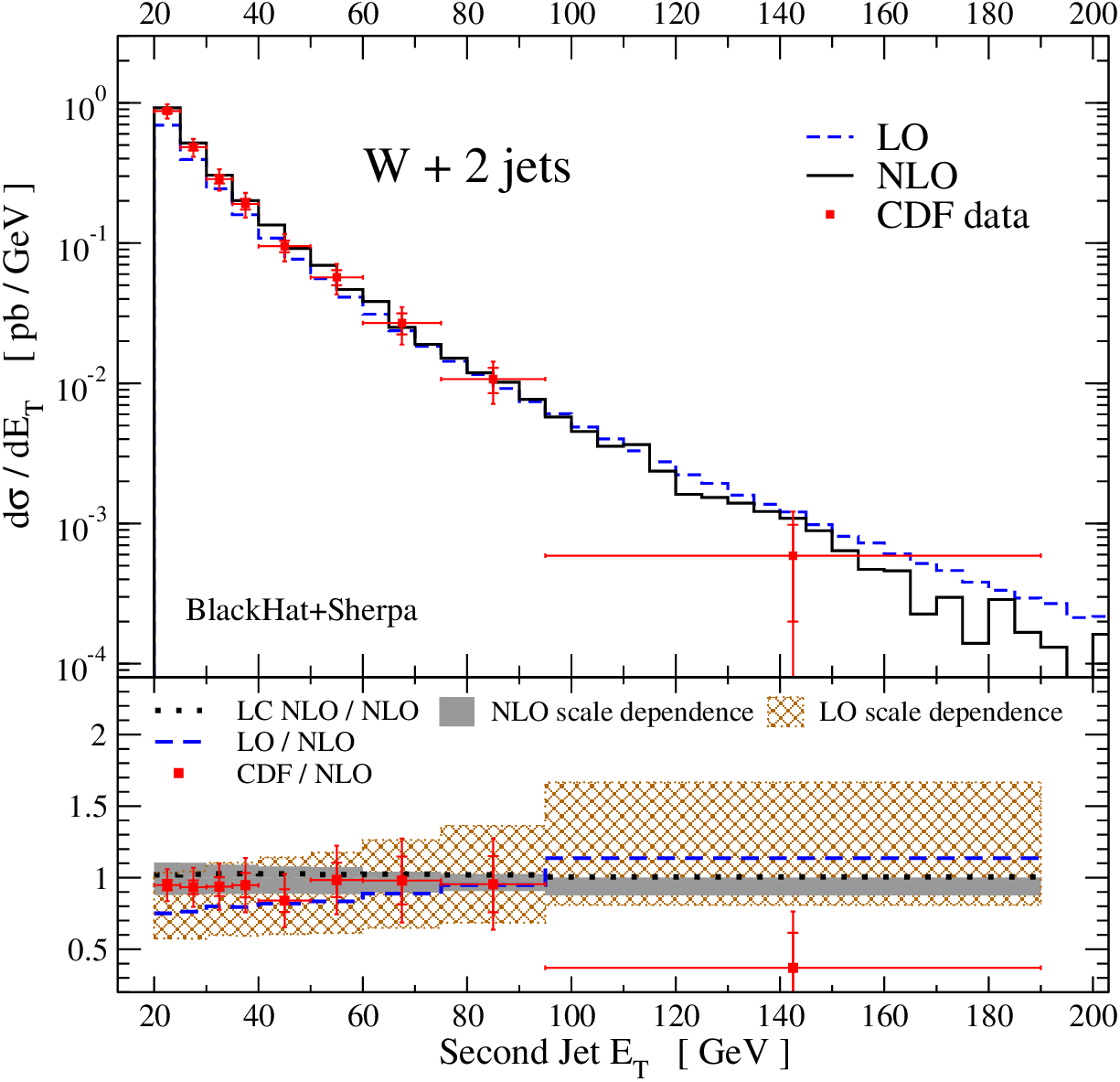}

\caption{NLO Differential cross section $d\sigma(W\rightarrow e\nu+\ge n-jets)/dE_T^{n{\rm th-jet}}$ for $n=1,2$ compared with the measured cross section. The upper panels contain the LO and NLO parton level predictions and the CDF data points with their statistical and total uncertainties represented by the inner and outer error bars respectively. The distributions normalized by the NLO prediction are shown in the lower panels. The scale uncertainty of the different predictions is represented by the shaded grey (NLO) and orange (LO) bands. The dotted black line represents our leading color approximation.\label{fig:W1W2}}
\end{figure}
\begin{table}
\begin{center}
\begin{tabular}{|c||c|c|c|}
\hline
number of jets  & CDF &  LC NLO & NLO  \\
\hline
1  & $\; 53.5 \pm 5.6 \;$ & $\; 58.3^{+4.6}_{-4.6} \;$ & 
            $\;  57.8^{+4.4}_{-4.0} \;$ \\
\hline
2  & $6.8 \pm 1.1$  & $ 7.81^{+0.54}_{-0.91}$ &
            $7.62^{+0.62}_{-0.86} $  \\
\hline
3 &  $0.84\pm 0.24$  & $\;0.908^{+0.044}_{-0.142} \;$ & ---  \\
\hline 
\end{tabular} 
\end{center}
\caption{
Comparison of the total cross sections in pb for $W\,+\,$n~jets with
 $E_T^{n\rm th\hbox{-}jet} > 25$ GeV from CDF to NLO QCD. 
For 1 and 2 jets the cross sections with and without LC approximation are displayed to show the quality of the approximation. For the three jets result, only the LC NLO result is currently available, but we expect a similarly small deviation for the full NLO result. 
The experimental statistical, systematic and luminosity uncertainties have been combined for the CDF results. \label{Table} 
}
\end{table}

We compare the NLO prediction for $W\,+\,$1,2,3~jets with data from the
CDF experiment~\cite{WCDF} at the Tevatron. For the analysis we have used 
the same cuts as in the CDF analysis with the SISCone~\cite{SISCONE} jet 
algorithm instead of the JETCLU~\cite{JETCLU} cone algorithm used by CDF, 
as the latter is not infrared safe. We set an event-by-event
renormalization and factorization scale according to $\mu=\sqrt{m_W^2+P_T^2(W)}$. 
  
\begin{figure}
\includegraphics[scale=0.27]{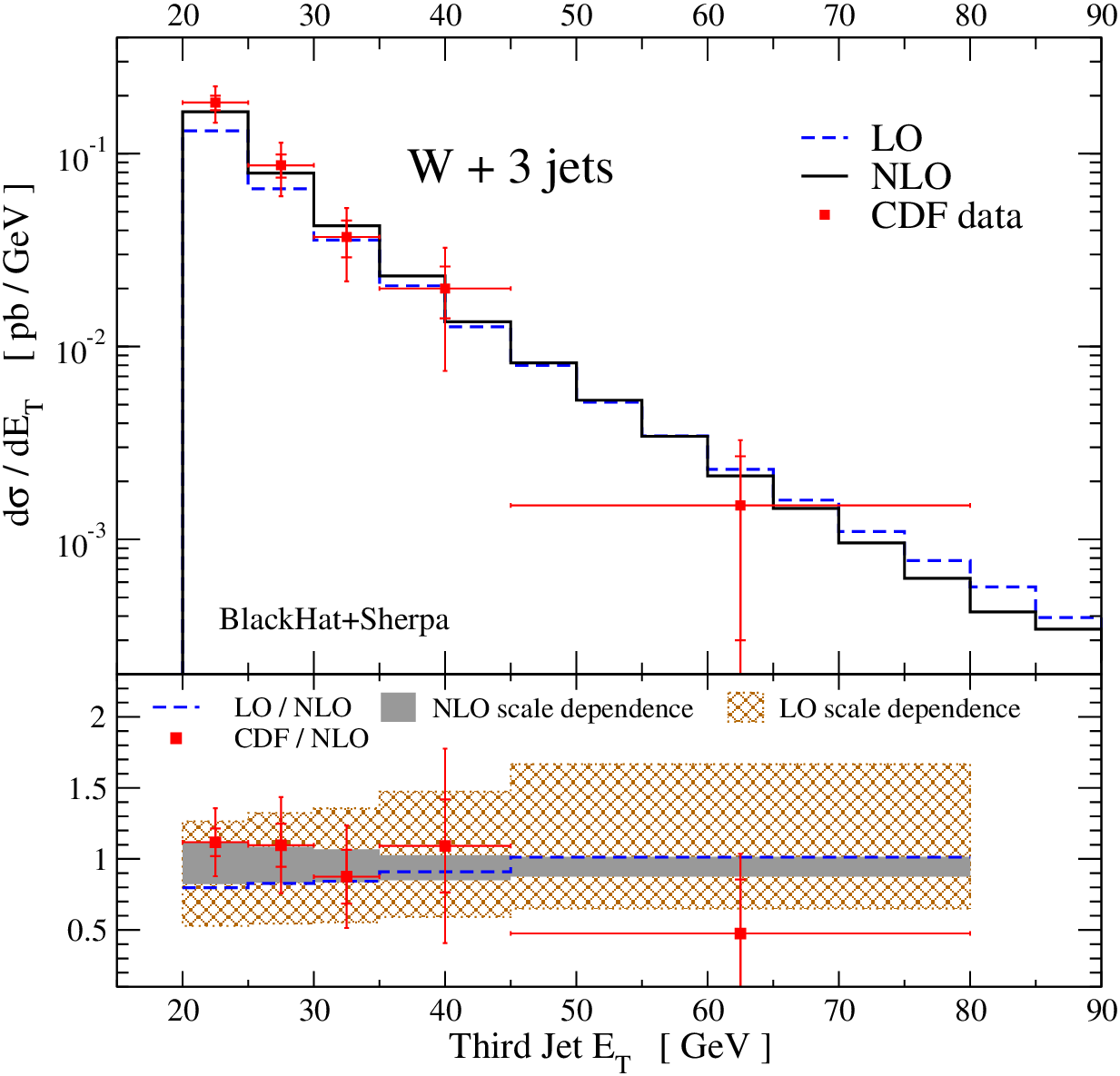}
\rule{1cm}{0cm}
\includegraphics[scale=0.27]{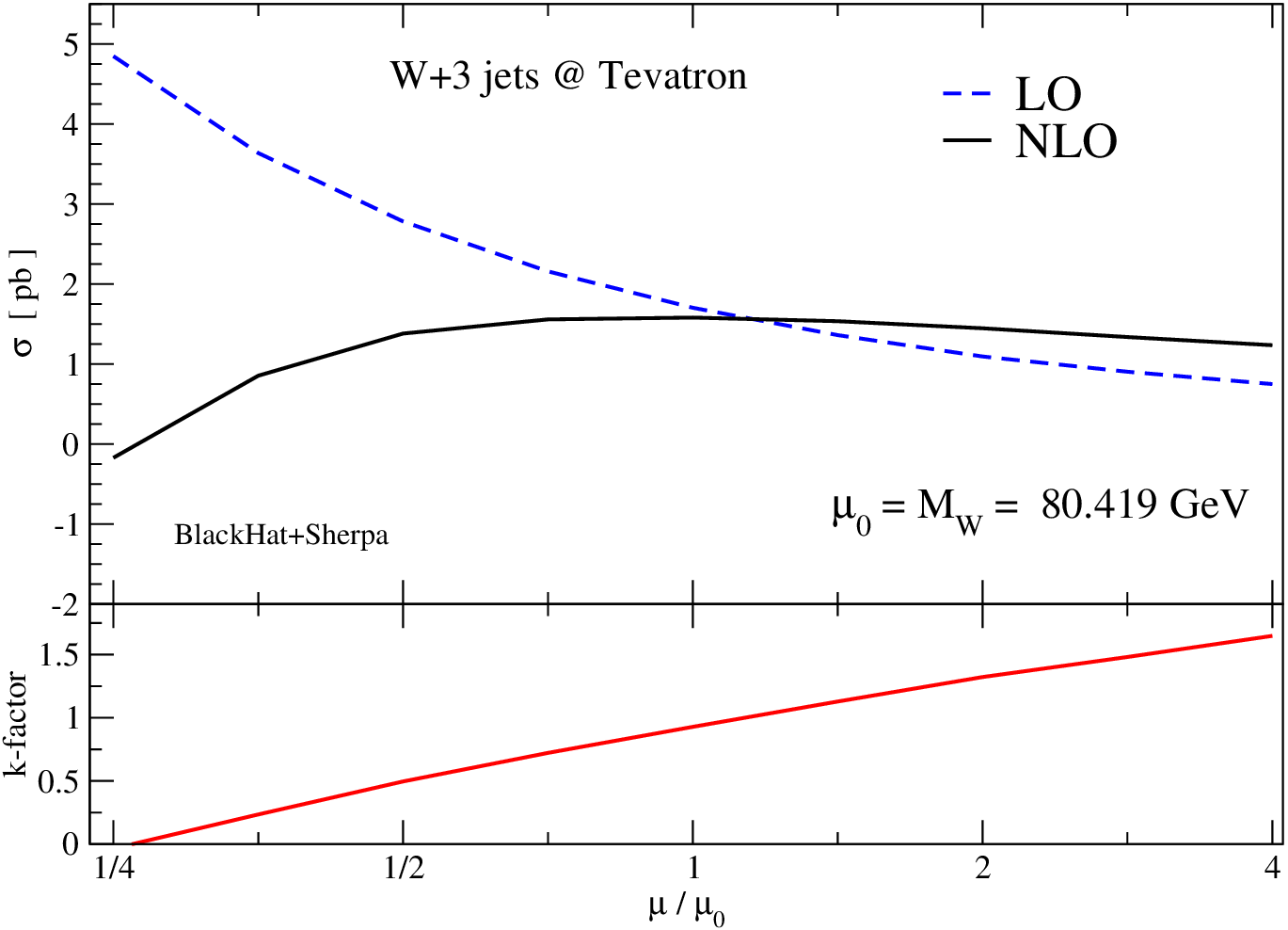}

\caption{The left panel shows the NLO Differential cross section
  $d\sigma(W\rightarrow e\nu+\ge 3-jets)/dE_T^{3{\rm rd-jet}}$  compared
  with the measured cross section. Its upper part shows the LO and NLO
  parton level predictions (within the LC approximation for the virtual
  part) and the CDF data points with their statistical and total
  uncertainties represented by the inner and outer error bars
  respectively. The normalized distributions are shown in the lower part
  of the left panel. The scale uncertainties are represented by the shaded
  grey (NLO) and orange (LO) bands. The right panel shows the scale
  dependence of the total cross section on the renormalization and
  factorization scale $\mu$,
  taken equal and varied between 1/4 and 4 times the mass of the $W$ boson.\label{fig:W3}}
\end{figure}

In Figure \ref{fig:W1W2} we present the transverse energy distribution 
of the $n$-th jet for $W\,+\,$1,2~jets jets production. The NLO result 
agrees with the previously available results from MCFM~\cite{MCFM}.
The lower part of these plots shows the reduced scale dependence of
the NLO prediction. The dotted black line demonstrates the validity of
our LC approximation across the whole $E_T$ range.
The plot in Figure \ref{fig:W3} shows a good agreement between the NLO
prediction (within our LC approximation) for the $E_T$ distribution of the third jet and the experimental data. The second plot of Figure 2 displays the improvement of the scale dependence of the cross section when NLO corrections are added. More details on the setup of our analysis can be found elsewhere~\cite{W3jets}. 

\section{$\boldmath{W}$+3 jets at the LHC}\label{sec:LHC}

\begin{figure}
\includegraphics[scale=0.27]{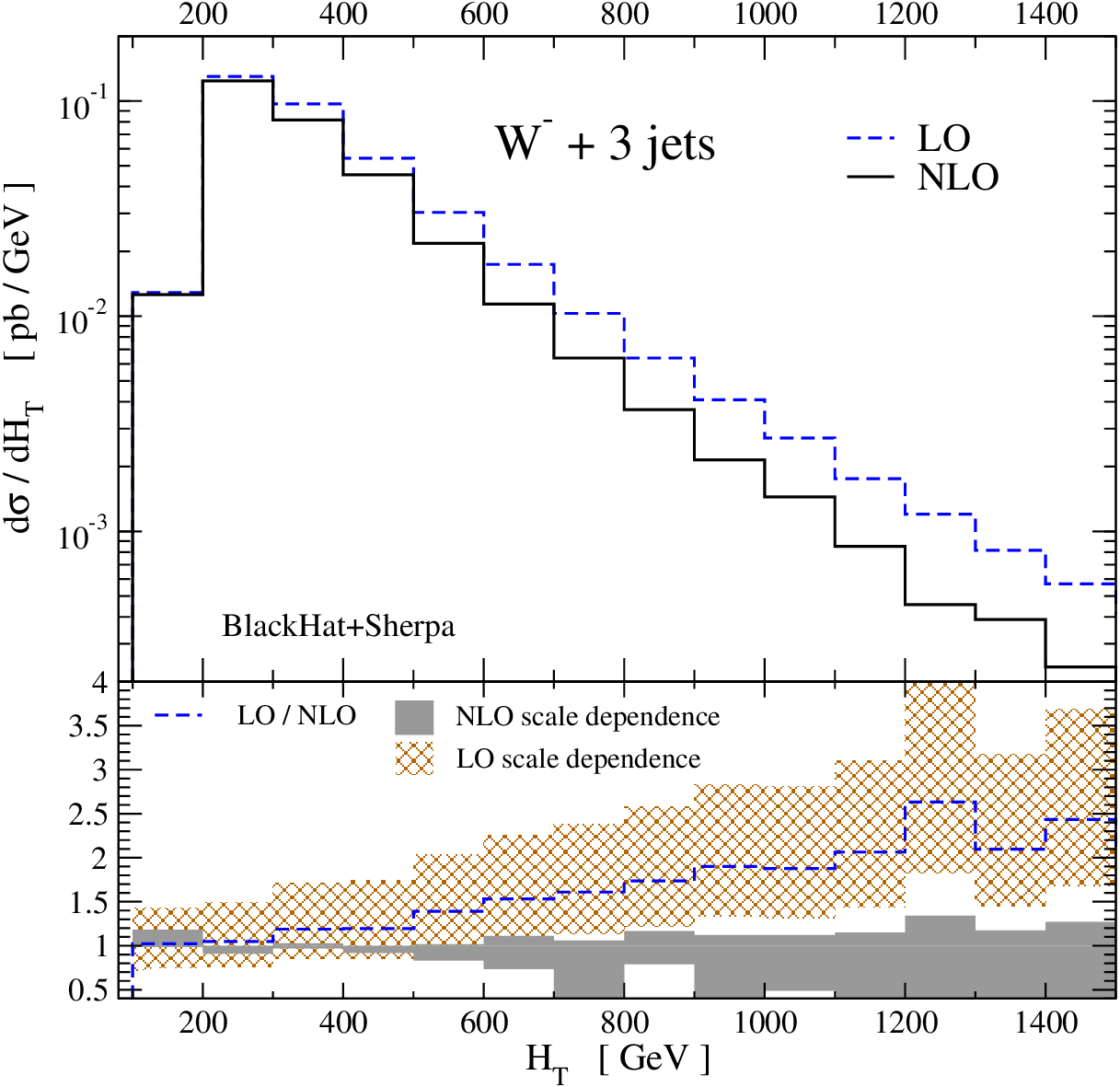}
\rule{1cm}{0cm}
\includegraphics[scale=0.27]{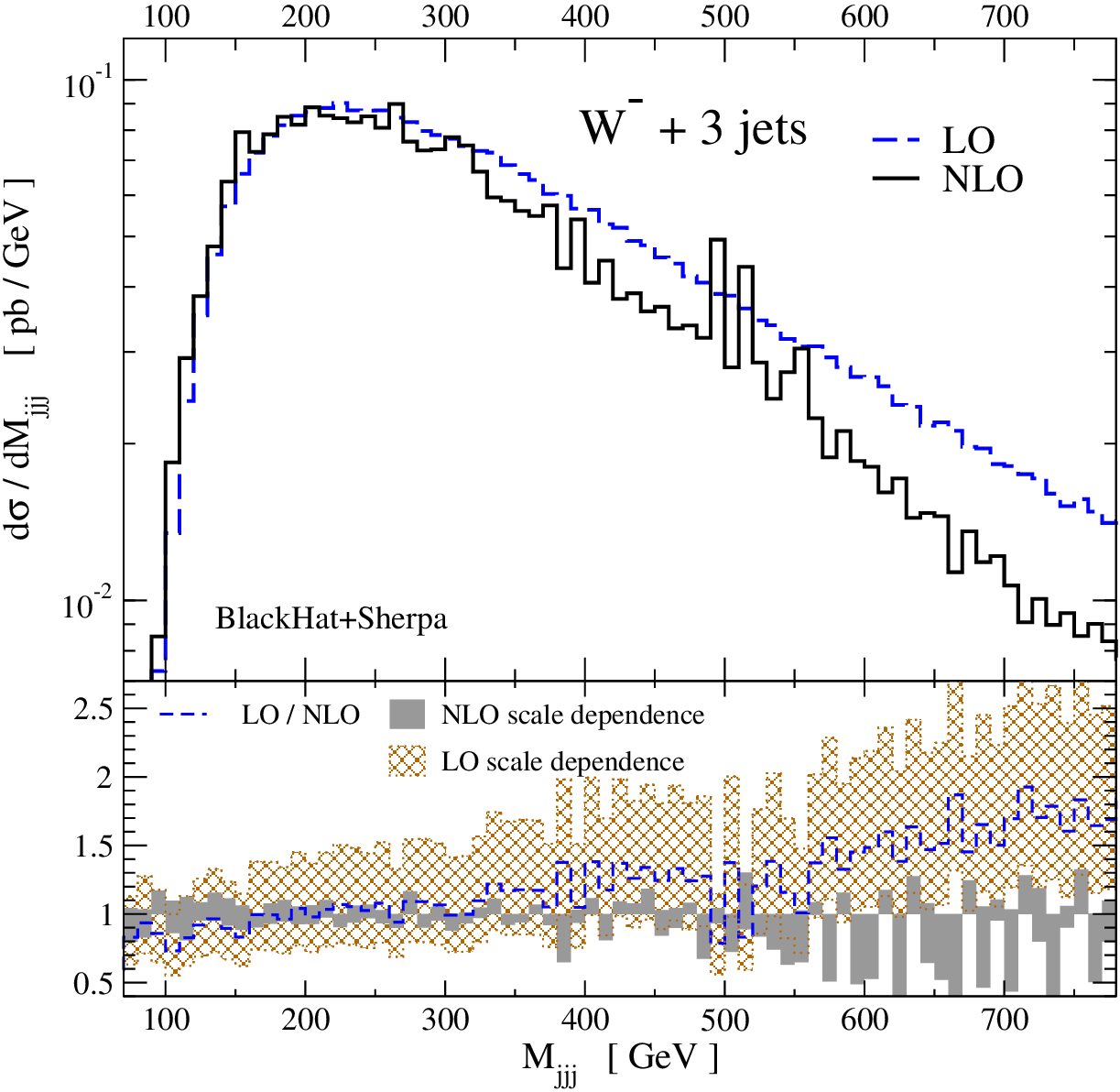}

\caption{ The NLO prediction for the scalar transverse energy sum $H_T$
 (left panel) and the three-jet mass $M_{\rm jjj}$ (right panel) 
 compared to the LO prediction. The LC approximation has been used
 for the virtual contribution to the NLO result. The scale uncertainties are represented by the bands in the lower parts of the plots.\label{fig:LHCW3}}
\end{figure}

We repeated the same analysis for the LHC with a
center-of-mass energy of 14~TeV. For this analysis,
we chose cuts suggested by the ATLAS and CMS technical design reports: 
$E_T^{e} > 20$ GeV,
$|\eta^e| < 2.5$, 
\rule{0.15cm}{0cm} $\not{\rule{-0.15cm}{0cm}E_T} > 30$ GeV, 
$M_T^W > 20$ GeV, and 
$E_T^{\rm jet} > 30$ GeV. 
Here the $E_T^i{}$s are transverse energies,
\rule{0.15cm}{0cm} $\not{\rule{-0.15cm}{0cm}E_T}$ 
is the missing transverse energy,
$M_T^W$ the transverse mass of the $e \nu$ pair, and $\eta$ the
pseudorapidity.  The $E_T$-ordered jets are required to have a rapidity in the range
$|\eta| < 3$. We used SISCone~\cite{SISCONE} with $R=0.4$. 
Figure \ref{fig:LHCW3} shows the distributions in the scalar transverse energy sum $H_T$
and in the three-jet mass 
$M_{\rm jjj}=\sqrt{\left(k_{j1}+k_{j2}+k_{j3}\right)^2}$
for $W^-\,+\,$3~jets at the LHC.

\section*{Acknowledgments}
This research was supported by the US
Department of Energy under contracts DE--FG03--91ER40662,
DE--AC02--76SF00515 and DE--FC02--94ER40818.  DAK's research is
supported by the Agence Nationale de la Recherche of France under grant
ANR--05--BLAN--0073--01, and by the European Research Council under
Advanced Investigator Grant ERC--AdG--228301.  This research used
resources of Academic Technology Services at UCLA and of the National
Energy Research Scientific Computing Center, which is supported by the
Office of Science of the U.S. Department of Energy under Contract
No. DE-AC02-05CH11231.

\end{document}